\documentclass[journal]{IEEEtran}

\usepackage{graphicx}

\usepackage[caption=false]{subfig}

\usepackage{multirow}
\usepackage{array}
\usepackage{algpseudocode}
\usepackage[table,xcdraw]{xcolor}
\usepackage{algorithm}
\usepackage{amsmath}
\usepackage{mathtools}
\usepackage{graphicx}
\usepackage{mwe}
\usepackage{adjustbox}
\usepackage{caption}
\usepackage{booktabs}
\usepackage[font={small,it}]{caption}
\usepackage{amssymb}
\usepackage{pifont}

\graphicspath {{./}}

\ifCLASSINFOpdf

\else

\fi

\begin{document}

\title{Scalable Attack-Resistant Obfuscation of Logic Circuits}

 \author{\IEEEauthorblockN{Abdulrahman Alaql  \ \ \ \  \ \ \ \  Swarup Bhunia}

 \IEEEauthorblockA{Dept. of ECE, University of Florida, Gainesville, FL 32608\\}
 \IEEEauthorblockA{ $alaql89@ufl.edu$, $swarup@ece.ufl.edu$}

 }

\maketitle

\begin{abstract}
Hardware IP protection has been one of the most critical areas of research in the past years. Recently, attacks on hardware IPs (such as reverse engineering or cloning) have evolved as attackers have developed sophisticated techniques. 
Therefore, hardware obfuscation has been introduced as a powerful tool to protect IPs against piracy attacks. 
However, many recent attempts to break existing obfuscation methods have been successful in unlocking the IP and restoring its functionality. 
In this paper, we propose SARO, a Scalable Attack-Resistant Obfuscation that provides a robust functional and structural design transformation process. 
SARO treats the target circuit as a graph, and performs a partitioning algorithm to produce a set of sub-graphs, then applies our novel Truth Table Transformation (T3) process to each partition. 
We also propose the $T3_{metric}$, which is developed to quantify the structural and functional design transformation level caused by the obfuscation process. 
We evaluate SARO on ISCAS85 and EPFL benchmarks, and provide full security and performance analysis of our proposed framework.

\end{abstract}

\IEEEpeerreviewmaketitle

\section{Introduction}

\begin{figure}[t]
\centering
\includegraphics[width=0.5\textwidth]{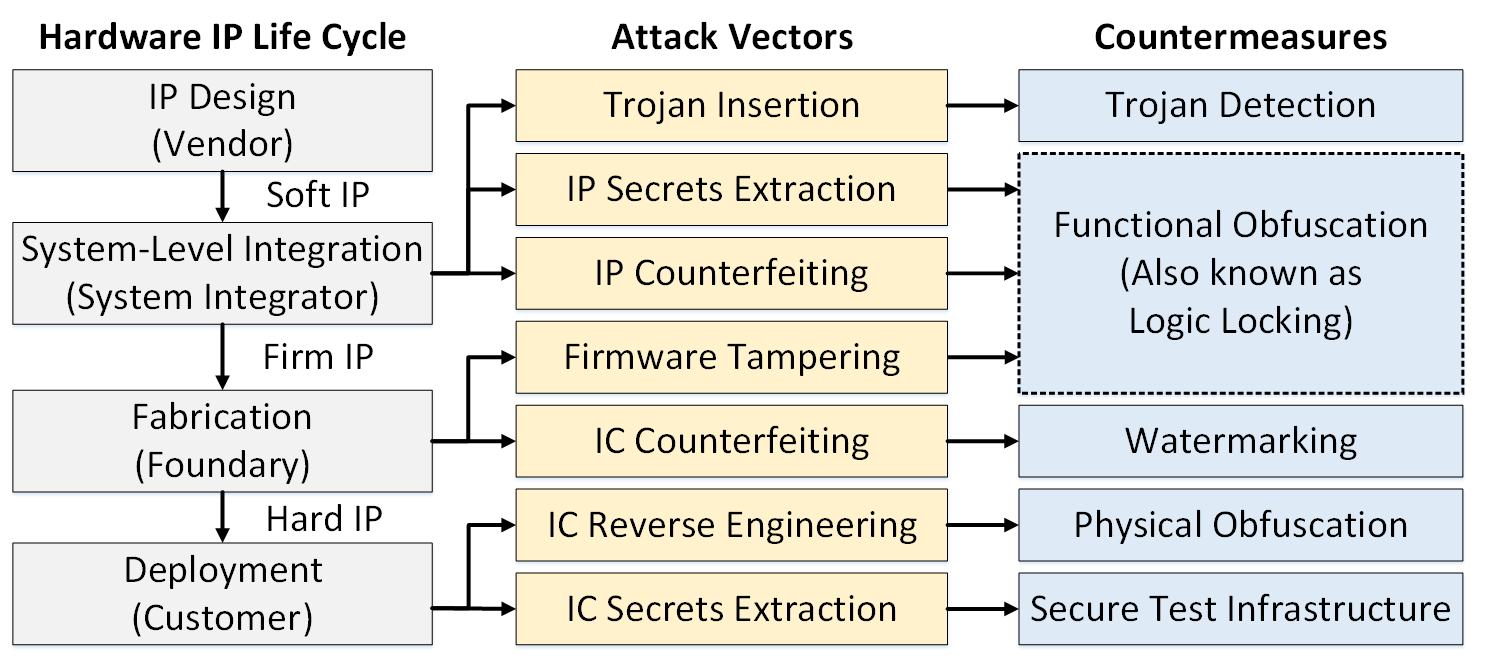}
\caption{The life-cycle of a hardware IP, untrusted parties may perform malicious activities to the IP at different phases. Since the focus of our approach is on soft IP protection, we assume that all other parties in the life-cycle of the IP are untrusted.}
\label{ven}
\end{figure}

Hardware IPs form a crucial component in the semiconductor industry. A single SoC comprises one or more IP blocks that are usually purchased from a third-party. Hardware IPs are typically available as RTL or gate-level netlists and can be integrated at any stage in the design flow. However, hardware IPs have been plagued by security concerns like IP piracy and counterfeiting in recent years. IP protection techniques can be categorized as authentication based, or obfuscation based. Authentication based techniques rely on the insertion of a unique signature (watermark) to prove ownership of the IP. Obfuscation techniques, on the other hand, rely on preventing the attacker from understanding the design-intent thereby preventing an unauthorized third-party from gaining access to, or replicating the design.
The IP life cycle is shown in Fig. \ref{ven} alongside some attack vectors and their possible countermeasures. 

Obfuscation techniques have significantly been evolving over the past years, where some are applied to the early stages of the IP's life cycle (functional), others are applied in the fabrication stage (physical) \cite{chakraborty2008hardware}. 
Physical obfuscation techniques are post-silicon measures that are applied to the basic structure of logic elements. By adding dummy contacts, these logic elements are not easily distinguishable, which adds a layer of protection against physical attacks. Functional obfuscation techniques (also known as logic locking) are applied to Register-Transfer Level (RTL) or gate-level netlists.
An example of functional obfuscation and a simple overview of the locking process are shown in Fig. \ref{obfus_history}. 
Various attacks have been proposed to break these obfuscation techniques. Over the years, these attacks have grown in sophistication as the attackers utilize information gleaned from every stage in the design process in order to bypass existing countermeasures.  

Unfortunately, most obfuscation techniques are developed with a singular goal of mitigating one type of attacks, while leaving the IP vulnerable to unconsidered and unexplored attacks.
This philosophy has caused a critical limitation, where existing obfuscation techniques lack a true design transformation.
For instance, some techniques focus on adding a single monolithic module at the circuit's output, leaving the rest of the design unchanged.
In other words, most existing techniques have failed to address the three essential properties of hardware obfuscation, which are the functional alteration, the structural alteration, and the scalability to large designs.

In this paper, we propose \textbf{SARO}, a \textbf{S}calable \textbf{A}ttack-\textbf{R}esistant \textbf{O}bfuscation of logic circuits. In this approach, we split the design into smaller partitions, and then perform a systemic Truth Table Transformation (T3) process to each partition.
The T3 process is an RTL-based locking mechanism implemented that is highly randomized in many aspects. The main benefit of obfuscating a small and local partition is to help in altering the structure of the netlist as well as the graphical representation of the overall design when applied to all partitions. 
Instead of inserting key-gates, we have developed SARO to perform a true design transformation to the entire circuit using functional and structural modifications that are able to completely hide the design intent of the circuit. 
SARO is extremely flexible, as it allows for choosing the size and number of partitions, as well as the type of transformation for each partition, which can help manage area and power overheads. 
Moreover, the proposed approach is shown to be scalable, as it requires polynomial time to perform the locking, which makes this approach suitable for obfuscating large IPs.
In addition to the proposed obfuscation technique, we present the $T3_{metric}$, a security metric that can quantify the level of structural and functional transformation of the locked circuit. This metric is scalable to large designs and can be applied to any obfuscation technique.
This paper makes the following major contributions:

\begin{figure}[!t]
\centering
\includegraphics[width=0.5\textwidth]{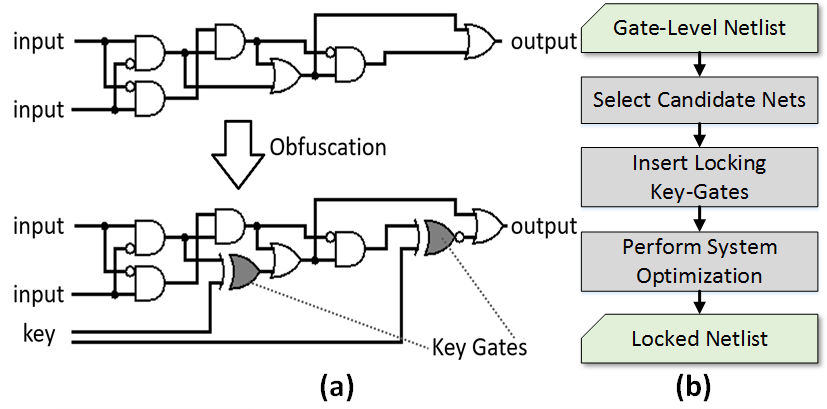}
\caption{An example of a locked circuit (a), using the process in (b).}
\label{obfus_history}
\end{figure}

\begin{itemize}

    \item  Analyzes the shortcomings of existing techniques, where security and performance issues deem these techniques unpractical or insecure.

    \item  Introduces SARO, a pre-silicon structural and functional hardware logic obfuscation, where the design is divided into small partitions, and each partition is obfuscated individually.
    This approach provides two main benefits, structural alteration of the circuit, and the ability to lock extremely large IPs, which makes the method scalable.

    \item  Performs a robust Truth Table Transformation (T3) process that offers a verity of locking mechanism functions and adds a key dependency to both the original and the added gates to form symmetric partition with hard-to-find dummy functions.

    \item  Introduces $T3_{metric}$, a pre-silicon obfuscation evaluation metric that can quantify the level of structural and functional alteration caused by the obfuscation process.
    
\end{itemize}

To the best of our knowledge, there has not been any work that focuses on providing a scalable framework to apply functional obfuscation with the focus on the structural alteration of the circuit. The rest of the paper consists of the following, 
Section II provides background regarding obfuscation techniques and attacks, and the issues of existing approaches. 
Section III introduces the methodology for the proposed obfuscation approach as well as the obfuscation metrics that can help evaluate the quality of the applied approach. 
Section IV discusses the implementation results of the approach in terms of runtime, overhead, and thorough security analysis of the proposed and the existing approaches. 
Finally, Section V concludes this study.

\subsubsection{\textbf{Node Selection Techniques}} 

Multiple node selection approaches have been developed; the following list overviews some of these approaches.

\paragraph{Random Insertion (RN)} 

A random node selection process is performed, where no heuristics are implemented. This approach is considered one of the original obfuscation techniques, which has been introduced in \cite{roy2008epic}. 
Although this method is scalable, there is no guarantee that the provided security for this approach is robust.

\section{Background and Motivation}

\subsection{\textbf{Background on Obfuscation Techniques}} 

\subsubsection{\textbf{Types of Obfuscation Key-Gates}} 

When considering functional obfuscation, there are different types of key-gates and dummy functions that can be added to selected nodes to be obfuscated. The following list shows the different types of obfuscation gates used for the selected nodes.

\paragraph{XOR/XNOR Gates} In this approach, an XOR/XNOR gate is inserted to the node and add a key signal to one of the inputs of the gate \cite{roy2008epic}. 

\paragraph{Multiplexers} A multiplexer is inserted in the selected node \cite{lee2015improving}, one of the inputs to the multiplexer is the original route, and the other inputs are for dummy functions and routes. 

\paragraph{Functional-Stripping Operations} Instead of placing a key-gate, a functional-stripping operation is performed to a set of nodes followed by a key-controlled functional restoring unit. The performed arithmetic operation is reversible, such that applying the correct key to the functional restoring unit restores the correct functionality. The best example of this approach is the Stripped-Functionality Logic Locking (SFLL), which has been introduced in \cite{yasin2017provably}.

\subsubsection{\textbf{Node Selection Techniques}} 

Different node selection approaches have been introduced in the literature; the following list elaborates on these approaches.

\paragraph{Random Insertion (RN)} 

The node selection in this approach is non-deterministic and random \cite{roy2008epic}. The process is quick and scalable as no analysis of the circuit is needed to insert key-gates. The generated list of selected nodes is changed every time the process is run.

\paragraph{Secure Logic Locking (SLL)} 

The node selection process introduced in \cite{yasin2015improving} is an iterative approach that avoids creating muting key-gates, which are defined as gates that are masked by other surrounding key-gates.

\paragraph{Logic Cone-Size (CS)} 

This selection process measures the logic cone-size of each node in the design \cite{lee2015improving}. The cone-size is calculated as the summation of the total number of nodes in the fan-in and fan-out of the considered node.

\paragraph{Controlability-Based Selection}

The node selection process in the technique analyzes the switching activity of the circuit and quantifies the probability for each node being switched based on a set of input patterns. This analysis is referred to as the controllability factor. The locking technique in \cite{dupuis2014novel} places key-gates in low controllability locations in order to prevent Trojan insertions.

\paragraph{Observability-Based Selection}

The node selection process analyzes the circuit for the probability of affecting the observed output, this analysis is referred to as the observability factor. 
The locking technique in \cite{alaql2019quality} selects nodes to obfuscate based on balanced observability values in order to control the output corruption level. This approach is designed to tolerate slight errors when the obfuscation key source is unreliable or noisy.

\subsection{\textbf{Attacks on Obfuscation}}

In this subsection, we discuss the existing attacks that can fully or partially break the locking of the obfuscated IPs. Based on our threat model, we focus on attacks that are applied to the locked gate-level netlist of the IP. 

\subsubsection{\textbf{Satisfiability Attack (SAT)}} 

SAT attack is a powerful tool that is applied to any logic-locked combinational circuit to extract the obfuscation key \cite{subramanyan2015evaluating}. To perform the attack, the obfuscated gate-level netlist, as well as an unlocked circuit, are required. The attack is applied by performing the SAT-solving algorithm; this algorithm applies different input patterns and key values, then compares the outputs of both the obfuscated and the unlocked circuits. The algorithm then uses these distinguishing input patterns (DIPs) to identify wrong key sub-sets, and the process iterates until the entire key-space is covered and the correct key is revealed. The attack is only applicable to purely combinational circuits, or sequential circuits with scan-chain capabilities. Which limits the number of IPs that SAT can be applied to. 
Another issue with the attack lays upon the applicability of SAT-complex designs (such as block-ciphers and multipliers), where the solvers are not able to reach a solution polynomial time.

\subsubsection{\textbf{Key Sensitization Attack (KSA)}} 

Key Sensitization Attack (KSA) is a technique that aims to force a primary output to leak a key-value \cite{yasin2015improving}. The attacker looks for a pattern that sensitizes a key input to the output. By identifying key-gates that can block the effect of other key-gates (muting gates), the attack can focus on sensitizing the key values of those gates. For the attack to be applied, an activated circuit is required alongside an obfuscated gate-level netlist. The major issue of this attack is scalability, where large and complex designs require a long time to sensitize the key values successfully. Additionally, there is no guarantee that all key values can be sensitized to a primary output of the design, especially in sequential circuits.

\subsubsection{\textbf{Hill-climbing Attack}} 

Hill-climbing is an iterative process that aims to maintain and improve the quality of the output. In the context of hardware obfuscation, this process can be used as an attack to obtain the obfuscation key \cite{plaza2015solving}. The hill-climbing attack requires a locked netlist, as well as a functional unlocked system. The golden test-patterns used for the system validation in the development stage can also be used instead of an unlocked system.

\subsubsection{\textbf{Template Matching Attack}} 

Template matching is a structural-based analysis attack that works in the manner of looking for recognizable patterns in the gate-level netlist \cite{li2013formal}. Attackers aim to use this approach to reverse engineer a system, where a high-level description of the system is publicly available, and recognizable patterns in the gate-level netlist are identified. The other use for this attack is to identify the critical parts (such as s-boxes in cryptographic modules), then modify the design to disable security modules, insert hardware Trojans \cite{chakraborty2009security}, or leak sensitive information to the primary output.

\subsubsection{\textbf{Structural Analysis using Machine Learning Attack (SAIL)}} 

SAIL is a machine learning structural analysis attack that aims to restore the obfuscated design into its original pre-synthesis form \cite{chakraborty2018sail}. The attack helps expose the intended key value for each key-gate inserted during the obfuscation process. SAIL is scalable and design-independent, as it can be applied to any obfuscated gate-level netlist that uses XOR/XNOR key-gates. In addition to the gate-level netlist, the attacker needs to obtain the obfuscation tool used on the target system. 
An enhancement to the attack algorithm, referred to as SURF, was introduced in \cite{chakraborty2019surf}. SURF proceeds the structural analysis by a hill-climbing process that is able to improve the accuracy of the attack.

\subsubsection{\textbf{Constant Propagation Attack (SWEEP)}} 

SWEEP is a structural analysis-based attack on logic locking \cite{alaql2019sweep}. The attack is performed by leveraging the information given by the synthesis tool to expose the correct key. The SWEEP attack analyzes each key input individually by replacing it with a logic zero in one instance, and a logic one in the other instance. In each instance, the modified locked netlist is synthesized, and the two synthesis reports are compared. The difference between the two reports (deltas) in terms of power, area, and critical path delay are observed and investigated for any correlation with the correct value of the key.

\subsubsection{\textbf{Attacks on SFLL}}

The functional analysis attack on logic locking (FALL) has been introduced in \cite{sirone2020functional} to break the obfuscation of the SFLL technique. The attack is specifically designed to identify the functionally stripped parts of the design, then it applies a functional analysis algorithm that can restore the original circuit. The attack was able to break SFLL with 90\% of successful attempts. 
Additionally, the attack in \cite{yang2019stripped} was able to break the obfuscation by introducing two algorithms that can find the protected patterns of SFLL.

\section{Scalable Attack-Resistant Obfuscation}

We propose SARO, a scalable attack-resistant obfuscation process that can be applied to protect systems against known attacks.  
In SARO, we provide a highly randomized design transformation solution that partitions the target circuit and locks each partition individually by employing our novel Truth Table Transformation (T3).
The two major benefits of this approach are 1) a strong structural alteration of the circuit is obtained, and 2) a framework that is scalable to large circuits.
Our approach is designed to be flexible to any system type, and to any key size. The power and area overheads are also controllable as the process accounts for the size of each partition and the total number of dummy functions to be inserted into the design. The following multi-stage process is shown in Algorithm \ref{obfus_flow}:

\begin{algorithm}[h]
\caption{Scalable Attack Resistant Obfuscation}
\label{obfus_flow}
\begin{algorithmic}[1]
\Procedure{SARO}{}\\
\textbf{Input:} $org\_netlist$: Design netlist to be obfuscated \\
\textbf{Input:} $k_{size}$: Size of the obfuscation key\\
\textbf{Input:} $aig\_flag$: AIG flag\\
\item $G_{org} \gets Hyper\_Graph\_Converter(org\_netlist)$
\item $Topo_{G} \gets Topological\_Sorting(G_{org})$
\item $partition_{size} \gets initial\_analysis(G_{org},k_{size})$ 
\item \textbf{if}  $aig\_flag$ \textbf{then} 
\item \hskip2em $G_{org} = AIG(G_{org})$ \Comment{optional}
\item $\mathbb{G}_{org}[p] = partition(G_{org},partition_{size})$ 

\item  \textbf{for} \text{$partition$ \textbf{in} $\mathbb{G}_{org}[p]$ }

\item \hskip4em $SARO\_netlist[p] = \textbf{T3}(partition)$ 

\item \textbf{end for}

\item  $locked\_netlist = merge(SARO\_netlist[p])$ \\

\textbf{Output:} $locked\_netlist$: Obfuscated netlist 

\EndProcedure
\end{algorithmic}
\end{algorithm}

\noindent The inputs to the algorithm are the original netlist $org\_netlist$, the size of the key, the value of the key (inserted or random), and overhead constraints (area, power, and timing). 
The netlist $org\_netlist$ of the target circuit is converted to its equivalent $G_{org}$.
The hypergraph is then topologically sorted, and each vertex is ranked based on its logic level. The topological dictionary $Topo_{G}$ is generated using the function introduced in \cite{cormen2009introduction}.
Then, we perform a partitioning algorithm as follows: $\mathbb{G}_{org}[p] = partition(G_{org},p)$, where $p$ is the number of partitions. 
Each partition in $\mathbb{G}_{org}[p]$ is obfuscated individually using our design transformation scheme. 
Each step in Algorithm \ref{obfus_flow} is designed to be highly randomized. For instance, after the partitioning is performed, a shuffling step (shown in line 12) is applied. This step ensures that non-deterministic design transformation solutions are offered.

\subsection {\textbf{Major Steps of SARO}}

\subsubsection {\textbf{Initial Analysis and Partition Size}}

In the initial analysis stage, the original netlist is compiled using a synthesis tool; the generated gate-level netlist is recompiled multiple times for optimization purposes. The optimized netlist $org\_netlist$ is considered the reference design, where area, power, and timing reports are used to measure the overheads for the obfuscated netlist produced at the end of the SARO process. Moreover, the size of partitions is specified in this stage based on the size of the circuit and the number of key-bits. The partition size is calculated using Eqn. \ref{eq_parti_size}.

\begin{equation} \label{eq_parti_size}
partition_{size} = round\_down\{  {Gates_{total} \over k_{size}} \} 
\end{equation}

\noindent where $Gates_{total}$ is the total number of gates (vertices), $k_{size}$ is the key size in bits, and $partition_{size}$ is the number of gates specified for each partition.

\subsubsection {\textbf{Universal Gate Transformation}}

This is an optional stage, where the design is transformed into its equivalent form of only AND gates and inverters, which is referred to as the And-Inverter Graph (AIG). These two gates are considered universal, as any logic function can be produced from these two. Although the generated netlist of this stage will be heavily unoptimized, applying this transform increases the ability to perform the design transformation to parts of the design that might not be accessible when using traditional libraries.

\subsubsection {\textbf{Hypergraph Partitioning}}

In this stage, the gate-level netlist is transformed into its equivalent hypergraph format. A hypergraph is a generalization of a graph in which a structure is represented with vertices and edges. Vertices are the basic elements (gates) in the hypergraph, while edges are the links (wires) that connect these vertices, hyperedges refer to edges that connect more than two vertices. 
The SARO tool assigns each gate and wire with a numerical value and adds gates to the vertices list, and wires to the edges list. The output of this stage is a hypergraph format file $G_{org}$ of the original netlist. The obtained hypergraph file is run through the partitioning process.
The used partitioning algorithm in SARO is called hMETIS \cite{karypis1999multilevel}, which is a partitioning framework that produces high-quality partitions in terms of the locality. The algorithm is designed to reduce the number of cut edges, where cut edges can be considered as the outputs of each partition. 
The partitioning framework uses a multi-level partitioning algorithm, which is described in the following 4-phase process:  

\begin{itemize}

\item \textbf{Coarsening Phase:} During this phase, the main hypergraph is split into a set of smaller subgraphs. This phase is applied to balance the number of hyperedges in each produced hypergraph. 

\item \textbf{Initial Partitioning Phase:} The initial partitioning phase computes the bisection of the coarsened hypergraphs generated in the first phase. Since the coarsened hypergraphs are relatively small (about 100 vertices each), the partitioning algorithm will not affect the runtime or the quality of partitions.

\item \textbf{Uncoarsening Phase:} In this phase, the partitions of the previous phase are used to reconstruct the main hypergraph. These partitions can be considered the vertices of a higher level graph.

\item \textbf{Refinement Phase:} A refinement algorithm is used to create the partitions for the main hypergraph. The objective of the refinement algorithm is to reduce the number of cuts in the main partitions. This phase produces the final list of partitions and their corresponding vertices.

\end{itemize}

The parameters for the partitioning tool are set based on the overhead constraints which dictate the average number of gates in each partition. The tool produces the file of the partitioned hypergraph $\mathbb{G}_{org}[p]$. The obfuscation tool then takes the partitions file, each line in this file represents a vertex, and the value in that line refers to the partition number to which that vertex is assigned to. The tool then reconstructs each partition in the gate-level and identifies the gates, inputs, wires, and outputs for each partition. Each partition is treated as a standalone design module.

\subsubsection {\textbf{Truth Table Transformation (T3)}}

In SARO, each partition is obfuscated by altering the overall functionality, this alteration is performed in RTL, where a key-based systemic T3 process is performed to genetically alter the partition's functionality.
Moreover, this transformation aims to maximize randomness while maintaining a robust locking mechanism. 
This is essential to combat any attack that focuses on machine learning training sets to extract the keys (such as SAIL). Hence, we focus on providing a non-deterministic obfuscation approach that is hard to predict.
Given the $i^{th}$ partition $P^i (I^i[n])=O^i[m]$, where $I^i$ is the local input bus to the partition with the size of $n$ bits, and $O^i$ is the local output bus of $P^i$ with the size of $m$ bits. 
For each output bit $t$ in $P^i$, where $0 \leq t < m$, the Boolean function $f^i_t$ can be expressed as $O^i_t = f^i_t (I^i_0,I^i_1,...,I^i_{n-1})$.
When performing the T3 process, the Boolean function $f^i_t (I^i) = O^i_t$ is transformed to $f'^i_t (I^i,K^i,D^i) = O'^i_t$, with the correct key being $K$, and the wrong key being $\overline{K}$.

Dummy inputs $D^i$ are taken from other partitions in the circuit, where the algorithm in SARO aims to take dummy inputs from other partitions in the same logic depth. This way a grid-like shape between partitions is formed, which causes a high graphical alteration of the circuit.
One challenging aspect of choosing dummy inputs is the accidental formation of combinational loops. These loops are formed when a combinational logic chain drives itself. In other words, if the output of a set of gates in a chain is connected to the input of the chain, then an oscillation behavior is formed. Good design practices encourage avoiding combinational loops, as they cause rapid switching and inaccurate timing analysis of the circuit.
In SARO, we avoid forming these loops by only choosing dummy inputs that do not exist in the fan-out cone of the output $O^i_t$, which can help to avoid these loops.
Moreover, we perform a topological ordering analysis of the circuit, where each node is ranked based on its logic depth from primary ports and memory elements (flip-flops).
Candidate dummy wires are selected to be less than or equal the lowest-ranked partition inputs $I^i[n]$, 
This process provides two benefits, firstly, it helps eliminate the formation of accidental combinational loops during the T3 process, and secondly, it reduces the chance of lengthening the critical path of the circuit, which is a critical factor to preserve the circuit's performance. 
The T3 subroutine process outlined in Algorithm \ref{algo_t3} shows how the transformation for partition $P^i$ is performed.

\begin{algorithm}[h]
\caption{Truth Table Transformation (T3) Process}
\label{algo_t3}
\begin{algorithmic}[1]
\Procedure{T3}{}\\
\textbf{Input:} $P^i$: Target Partition $i$ \\
\textbf{Input:} $Topo_{G}$: Topologically Sorted Vertices Dictionary
\item $T3\_flag \gets partition\_evaluator(P^i)$ \Comment{optional} 
\item \textbf{if}  $T3\_flag$ \textbf{then} 
\item \hskip2em return $P^i$

\item \textbf{else} 

\item \hskip2em $P'^i \gets$ \{\} \Comment{initialize transformed partition} 
\item \hskip2em $f^i_t \gets function\_extractor(P^i)$
\item \hskip2em $D^i \gets dummy\_inputs\_assignment(Topo_{G},P^i)$

\item  \hskip2em \textbf{for} $q = 1$ \textbf{to}   $t$ \Comment{partition output location} 

\item \hskip4em $f'^i_q (I^i,K^i,D^i) \gets Transform(f^i_t (I^i))$
\item \hskip4em $v'^i_q \gets RTL\_generator(f'^i_q)$ 
\item \hskip4em $P'^i \gets append(v'^i_q)$

\item \hskip2em return $P'^i$

\EndProcedure
\end{algorithmic}
\end{algorithm}

\begin{figure}[h]
\centering
\includegraphics[width=0.5\textwidth]{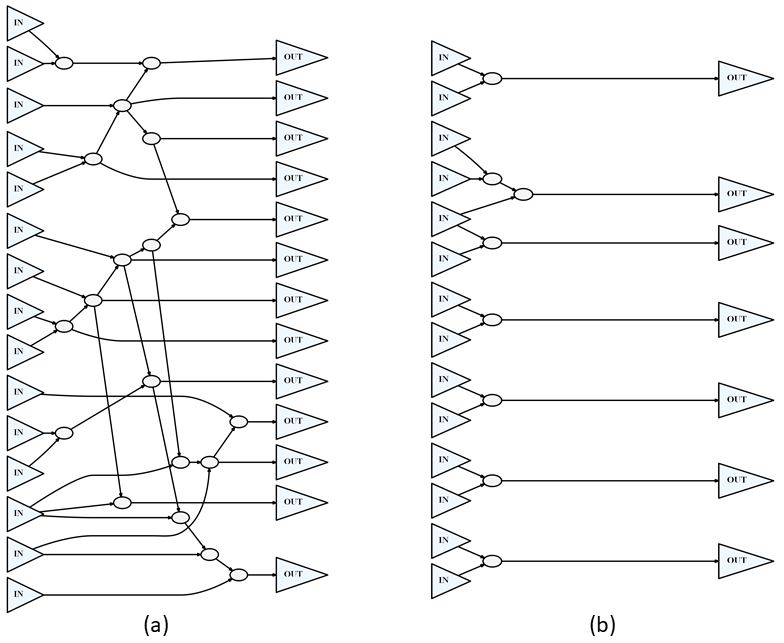}
\caption{(a) is an example of a partition that is suitable for the T3 process. While the partition in (b) is an example of a partition with disconnected fan-in cones and low logic depth, which may introduce large overhead when performing T3. }
\label{partitions_eval}
\end{figure}

The T3 process takes both the partition $P^i$ and the topologically sorted dictionary $Topo_{G}$ as input. 
The process starts with evaluating the partition topologically, where two major properties are investigated. 
Firstly, the fan-in cone for each output is analyzed, where cones that cover more inputs are more suitable for T3 than the others. 
Secondly, a logic depth analysis for the entire partition is performed, where the maximum logic depth is reported.
Based on these two properties, the T3 process is either performed or skipped.
The user can adjust the requirements for allowing T3 to be applied, or skip the evaluation altogether.
Two partition examples that illustrate suitable and non-suitable partitions for T3 are shown in Fig. \ref{partitions_eval}.

The example in Fig. \ref{partitions_eval}.(a) shows a well-connected partition, where fan-in cones cover a large number of inputs. Additionally, the logic depth for this example is high, which allows for a more successful design transformation process.
On the other hand, the example in Fig. \ref{partitions_eval}.(b) shows a disconnected partition, these typed of partitions are generated due to the limitation of the partitioning algorithm, and the type of target design. In this example, most fan-in cones are completely disconnected, and the logic depth of the overall partition is too low.
Although T3 can still transform this partition and introduce a robust design modification, it may introduce additional overhead.
When the $T3\_flag$ is true, a functional representation $f^i_t$ for each output $O^i_t$ in the partition is then generated by extracting all logic components controlling the considered output in the partition.

Dummy inputs $D^i$ are also generated by selecting wires in the design that are less than or equal to the lowest-ranked vertex in $f^i_t$.
The number and location of selected wires of $D^i$ are randomized, ranging from $0$ to $n$ the size of the original partition input $I^i[n]$.
The transform process is then performed to $f^i_q$, where one of the candidate operations is added to the function to form $f'^i_q$.
After the function is transformed, a separate file of the modified function $f'^i_q$ (expressed in Verilog RTL) is generated.
The algorithm concludes when all functions are transformed, and the modified partition $P'^i$ is obtained by groping all of the created $f'^i_q$ functions.
The following list in Table \ref{case_table} different types of functional and structural altering operations that are added to the CASE statement:

\begin{table}[h]
\centering
\caption{A Summary of Transformation Operations used in T3}
\label{case_table}
\resizebox{\linewidth}{!}{%
\begin{tabular}{|c|c|c|}
\hline
\textbf{Modification Type} & \textbf{Correct Key Applied} & \textbf{Wrong Key Applied} \\ \hline

Shuffled Outputs & $f'^i_t (I^i,K,D^i) = O^i_{t}$      &   $f'^i_t (I^i,\overline{K},D^i) = O^i_{t-1}$   \\ \hline
Arithmetic Operation & $f'^i_t (I^i,K,D^i) = O^i_{t}$    &   $f'^i_t (I^i,\overline{K},D^i) = O^i_{t} - K$      \\ \hline
Inverted Outputs & $f'^i_t (I^i,K,D^i) = O^i_{t}$       &   $f'^i_t (I^i,\overline{K},D^i) =  inv(O^i_{t})$   \\ \hline
Dummy Substitution    & $f'^i_t (I^i,K,D^i) = O^i_{t}$       &   $f'^i_t (I^i,\overline{K},D^i) = f^i_{t}(D^i)$  \\ \hline
Random Function    & $f'^i_t (I^i,K,D^i) = O^i_{t}$       &   $f'^i_t (I^i,\overline{K},D^i) = R^i_{t}(I^i,\overline{K},D^i)$  \\ \hline
\end{tabular}}
\end{table}

The listed transformation operations offer a wide range of modification options to choose from. For example, shuffled outputs can be implemented using different output sequences. 
On the other hand, arithmetic operations can corrupt the output by performing a key-based randomly selected function.
Simple operations such as inverting the outputs are also used, as well as introducing a random function $R^i_{t}$ that provides a unique behavior.
The application of these modification operations results in a strong functional and topological impact on the partition.

\subsection {\textbf{Complexity Analysis for SARO's T3}}

The Boolean function $f^i_t$ can be represented as a look-up table consisting of $n$ select-lines, one output, and $2^n$ configuration entries.
The entries of the $f^i_t$ look-up table express the behavior of the function across all possible combinations of $I^i$.
However, when performing SARO's design transformation to $P_i$, the Boolean function $f^i_t$ representing the output $O^i_t$ is modified to  $O'^i_t = f'^i_t (I^i_0,I^i_1,...,I^i_{d-1},D^i_0,D^i_1,...,D^i_{d-1},K^i_0,K^i_1,...,K^i_{k-1})$, where $D^i$ is the dummy input bus with the size of $d$ bits, and $K^i$ is the key input bus with the size of $k$ bits.

The beauty of incorporating dummy inputs alongside the added key inputs is to extend the size of the truth table of $f'^i_t$, which in turn allows for a wide range of corrupted functions to be incorporated, which suggests that the number of possible functions increases exponentially as the total number of inputs increase.
The total number of possible functions can be expressed in Eqn. \ref{eq_lut2} and \ref{eq_lut}:
\begin{equation} \label{eq_lut2}
E = (2^{n+k+d}-2^{n+d})
\end{equation}
\begin{equation} \label{eq_lut}
F_{count} = 2^{{E}}
\end{equation}
\noindent where $E$ is the number of entry locations for the corrupted functions, $n$, $k$, and $d$ are the input, key, and dummy input bus sizes respectively. The number of entries $E$ is computed by excluding the entries allocated for the original function, which is shown in the subtracted in the equation. 
$F_{count}$ is computed as $2^E$, which is a high-order function that represents the range of possible configurations that SARO can implement in the locked mode. 
Since the original partition inputs $I^i$, and the number of allocated key inputs to the partition $K^i$ are usually fixed, SARO is able to control the size of $D^i$ in order to expand the choices for corrupted functions.
A simple analysis in Table \ref{lut_count} for a 3-input partition shows the total number of possible corrupted functions $F_{count}$:

\begin{table}[h]    
\centering
\caption{A Summary of Attacks on Obfuscated Gate-level Netlists}
\label{lut_count}
\resizebox{\linewidth}{!}{%
\begin{tabular}{cccccc}
\hline

\multicolumn{3}{c}{\textbf{Input Size}} &\multirow{2}{*}{ \textbf{Output Size}} & \multirow{2}{*}{\textbf{$E$}}  & \multirow{2}{*}{\textbf{$F_{count}$}}  \\ \cline{1-3}
Original ($n$) &  Key ($k$) & Dummy ($d$)&  &    \\  \hline
3 & 1 & 0 &  1    &   8   &   256       \\ 
3 & 1 & 1 &  1    &   16  &   6.55e+5   \\ 
3 & 1 & 2 &  1    &   32  &   4.29e+10  \\ 
3 & 2 & 0 &  1    &   24  &   1.67e+7   \\ 
3 & 2 & 1 &  1    &   48  &   2.81e+14  \\
3 & 2 & 2 &  1    &   96  &   7.92e+28  \\ 
3 & 3 & 0 &  1    &   56  &   7.20e+16  \\
3 & 3 & 1 &  1    &   112 &   5.19e+33  \\ 
3 & 3 & 2 &  1    &   224 &   2.69e+67  \\ 
\hline
\end{tabular}}
\end{table}

\begin{figure}[h]
\centering
\includegraphics[width=0.5\textwidth]{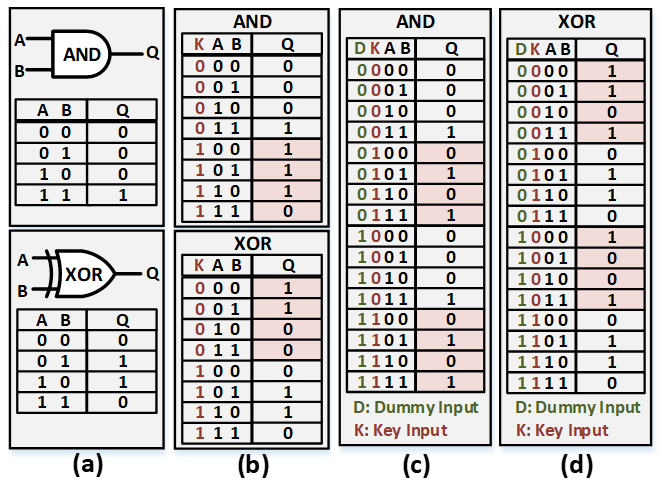}
\caption{(a) Examples AND and XOR functions represented in look-up table formats, (b) shows the traditional obfuscation of the functions, and (c) and (d) show the proposed design transformation processes for AND and XOR functions respectively.}
\label{lut}
\end{figure}

The computed values in Table \ref{lut_count} suggest that introducing more dummy inputs $d$ allows for an extremely wide range of choices for the corrupted function to implement. 
The major benefit of introducing this level of flexibility is to allow for a robust structural and functional design transformation process.
Fig.\ref{lut} provides two examples that show how the T3 process offers a higher number of corrupted functions to choose from.

Fig.\ref{lut}.(a) shows AND and XOR functions represented in a look-up table format, where input values are highlighted in green, and their corresponding outputs are highlighted in blue. 
Fig.\ref{lut}.(b) shows both functions being obfuscated using traditional key-gate insertion, where the key input (highlighted in red) determines the behavior of the function and outputs highlighted in red are for the locked mode.
When using SARO's T3 process, additional dummy inputs (highlighted in green) are utilized to widen the range of possible corrupted functions.
Fig.\ref{lut}.(c) and (d) show how the design transformation process is applied to AND and XOR gates respectively, where the additional dummy inputs (digits highlighted in green) added a wide range of possibilities.

\subsection{\textbf{Obfuscation Evaluation Metrics}}

In this subsection, we use three metrics that can quantify the structural and functional impact caused by the obfuscation. These metrics can be used to evaluate the existing obfuscation schemes, as well as SARO.

\subsubsection{Scatter Index}
The first metric we use is the scatter index of key-entry nodes, which are defined as the first node that is connected to a key input.
We focus on these nodes because they determine the location at which the design modification is performed.
The proposed scatter metric measures how well-distributed key-entry nodes are in the obfuscated design. The index is calculated by performing a scoring process that evaluates each node in the design and determines whether the node is within an acceptable logic-level-distance from the closest inserted key-entry node. The process is described in Algorithm \ref{algo2}:

\begin{algorithm}[h] 
\caption{Scatter Index Evaluation}
\label{algo2}
\begin{algorithmic}[1] 
\Procedure{Scatter Index Scoring}{}\\
\textbf{Input:} $L_{depth}[n]$: Logic depth between node $n$ and the closest key-entry node \\
\textbf{Input:} $n$: Number of nodes\\
\textbf{Input:} $max_{depth}$: Maximum acceptable logic depth
$scatter\_score =0$

\item   \textbf{for} $i = 1$  \textbf{to} $n$

\If {$L_{depth}[i]  \leq   max_{depth}$}

$scatter\_score = scatter\_score+1$

\EndIf

\item   \textbf{end for} \\

$T_{index} = {scatter\_score \over n} \times 100\%$ \\
\textbf{Output:} $T_{index}$: Scatter index
\EndProcedure
\end{algorithmic}
\end{algorithm}

\noindent where $T_{index}$ is the scatter index, $n$ is the number of nodes in the design, and $L_{depth}[n]$ is the logic depth between node $n$ and the closest key-entry node. The parameter $max_{depth}$ is the maximum acceptable logic-level-distance between a node and the closest key-entry node; this parameter is set by the user to determine the acceptable level of distributed key-entry nodes.
The main two factors affecting $max_{depth}$ are the key size and the number of nodes $n$.
In our analysis, we set $max_{depth}={n \over{ k_{size}}}$. The function in Algorithm \ref{algo2} performs a scoring process that evaluates each node in the obfuscated design. For each node $i$, the variable $scatter\_score$ is incremented when the logic-depth-distance between it and the closest key-entry node is less than or equal to $max_{depth}$.

\subsubsection{Obfuscation Coverage Index}
The second metric we introduce is the obfuscation coverage, which is the ratio of gates affected by the obfuscation to the total number of gates in the design. This coverage metric will quantify the propagation of data corruption caused by locked key-entry nodes. The metric is measured by identifying all gates that are included in the fan-out logic cones of all key-entry nodes, and a percentage of coverage is calculated by Eqn. \ref{eq_cover}:

\begin{equation} \label{eq_cover}
C_{index} = { \sum_{k=1}^{i_{keys}} G_{covered} \over G_{total} } \times 100\%
\end{equation}

\noindent where $C_{index}$ is the coverage index, $i_{keys}$ is the number of key-entry nodes in the design, $G_{covered}$ is the number of gates inside the fan-out cones of key-entry node $k$, and $G_{total}$ is the number of gates in the design. Obfuscated designs with high coverage indexes make it difficult for attackers to analyze the behavior of the circuit when in locked mode and make the obfuscated function hard to identify.

\subsubsection{Formal Verification}

Formal verification is a process that compares two versions of a design and confirms if they are functionally equivalent. This process applies a set of input patterns that are selected to trigger unique and hard-to-reach areas of the design. These patterns are applied to both designs, and the outputs of the designs are compared. The test pattern passes the test if both designs provide an equivalent output, while unmatched outputs are considered a failure. Formal verification can be used to quantify the amount of functional change between the original and the obfuscated circuits. We use the formality index metric, which represents the percentage of corruption out of the total set of patterns applied by the formal verification tool.
Eqn. \ref{eq_hamm} and Eqn. \ref{eq_formal} show how the formality metric $F_{index}$ is calculated: 

\begin{equation} \label{eq_hamm}
Pattern_{match} ={ O_{size}-[Hamming(O_{obf},O_{golden})] \over O_{size}}
\end{equation}

\begin{equation} \label{eq_formal}
F_{index} = { \sum_{k=1}^{P_{total}} (1- |2*Pattern_{match}[k] - 1|) \over P_{total}} \times 100\%
\end{equation}

\noindent the matching rate $Pattern_{match}$ is used to examine each input-output pattern. The golden output $O_{golden}$ is obtained from the unlocked circuit, while the output for the obfuscated circuit $O_{obf}$ is obtained after a randomly selected key is applied. The operation $Hamming(O_{obf},O_{golden})$ performs the hamming distance calculation between the two obtained outputs, where the maximum Hamming distance value is $O_{size}$, which is the size of the output bus. The formality index $F_{index}$ is obtained by computing the sum of the mismatch rate, divided by the total number of applied input patterns ${P_{total}}$.

All three metrics are used to create the $T3_{metric}$, which quantifies the structural and functional design transformation level of the obfuscation process. Eqn. \ref{eq_s} shows how the $T3_{metric}$ metric is calculated:

\begin{equation} \label{eq_s}
T3_{metric} = [F_{index} +  C_{index} +  T_{index}] / 3 
\end{equation}

The indexes $T_{index}$, $C_{index}$, and $F_{index}$ are obtained from Algorithm \ref{algo2}, Equations \ref{eq_cover}, and \ref{eq_formal} respectively. Each of these indexes is equally weighted.

\section{Results and Analysis}

\begin{table*}[]
\centering
\caption{Comparison of Structural Differences between Various Obfuscation Techniques}
\label{struc}
\resizebox{\linewidth}{!}{%
\begin{tabular}{cccclccclccclccc}
\hline
\multicolumn{1}{l}{\multirow{3}{*}{Benchmark}} & \multicolumn{3}{c}{Cone-Size Based Obfuscation (CS)}              &    & \multicolumn{3}{c}{Secure Logic Locking (SLL)}                 &  & \multicolumn{3}{c}{Randomly Inserted (RN)}  &  & \multicolumn{3}{c}{\textbf{Proposed Approach (SARO)}}                                   \\  \cline{2-4} \cline{6-8} \cline{10-12} \cline{14-16}
\multicolumn{1}{l}{}                           & \begin{tabular}{@{}c@{}}Formality \\ Index (\%)\end{tabular}  & \begin{tabular}{@{}c@{}}Scatter \\ Index (\%)\end{tabular} & \begin{tabular}{@{}c@{}}Coverage \\ Index (\%) \end{tabular}  & & \begin{tabular}{@{}c@{}}Formality \\ Index (\%)\end{tabular}   & \begin{tabular}{@{}c@{}}Scatter \\ Index (\%)\end{tabular} & \begin{tabular}{@{}c@{}}Coverage \\ Index (\%) \end{tabular}  & & \begin{tabular}{@{}c@{}}Formality \\ Index (\%)\end{tabular}        & \begin{tabular}{@{}c@{}}Scatter \\ Index (\%)\end{tabular} & \begin{tabular}{@{}c@{}}Coverage \\ 
Index (\%) \end{tabular}     &       &     \begin{tabular}{@{}c@{}}Formality \\ Index (\%)\end{tabular}        &    \begin{tabular}{@{}c@{}}Scatter \\ Index (\%)\end{tabular}    &    \begin{tabular}{@{}c@{}}Coverage \\ Index (\%) \end{tabular}  \\ 

\hline

c432       &    47.2   &   13.4   &   72.4    & &   55.7   &   5.8   &   44.7   & &   47.5   &   6.6   &   22.5   &  &  75.5   &   93.5   &  83.5  \\
c499       &    65.3   &   14.3   &   65.7    & &   45.1   &   7.9   &   61.2   & &   44.3   &   8.7   &   28.6   &  &  78.3   &   94.5   &  77.4  \\
c880       &    49.3   &   17.8   &   81.2    & &   41.3   &   8.1   &   17.1   & &   27.5   &   13.8  &   31.8   &  &  82.2   &   94.0   &  66.9  \\
c1355      &    58.3   &   18.7   &   72.6    & &   22.8   &   10.8  &   19.6   & &   77.6   &   15.7  &   41.5   &  &  67.6   &   94.8   &  64.1  \\
c1908      &    55.6   &   15.6   &   64.2    & &   47.5   &   11.7  &   12.8   & &   57.2   &   12.1  &   27.5   &  &  100    &   93.7   &  71.5  \\
c2670      &    67.6   &   17.8   &   80.1    & &   39.6   &   14.8  &   33.6   & &   48.6   &   17.7  &   51.7   &  &  87.1   &   93.7   &  71.1  \\
c3540      &    64.3   &   19.5   &   79.7    & &   74.4   &   21.4  &   45.7   & &   66.7   &   22.3  &   28.6   &  &  86.3   &   93.3   &  69.5  \\
c5315      &    68.2   &   48.7   &   74.3    & &   65.8   &   20.4  &   28.9   & &   47.9   &   27.2  &   27.4   &  &  75.3   &   95.3   &  80.9  \\
c6288      &    61.2   &   41.8   &   68.4    & &   59.9   &   19.7  &   44.2   & &   51.7   &   19.3  &   22.2   &  &  90.6   &   96.1   &  91.3  \\
c7552      &    68.6   &   69.6   &   66.3    & &   45.2   &   27.5  &   47.8   & &   44.8   &   22.9  &   51.4   &  &  75.5   &   93.6   &  78.9  \\
ALU$^{*}$        &    N/A    &   N/A    &   N/A     & &   N/A    &   N/A   &   N/A    & &   34.6   &   27.1  &   48.3   &  &  68.3   &   99.1   &  81.2  \\
mem\_ctrl$^{*}$  &    N/A    &   N/A    &   N/A     & &   N/A    &   N/A   &   N/A    & &   41.6   &   34.8  &   36.6   &  &  71.6   &   93.6   &  79.3  \\
FIR$^{*}$        &    N/A    &   N/A    &   N/A     & &   N/A    &   N/A   &   N/A    & &   58.3    &   44.7  &   39.4   &  &  82.4    &   91.5   &  84.2  \\
\hline
\textbf{Average$^{**}$}   & \textbf{60.56}     & \textbf{27.72}     & \textbf{72.49}     & &  \textbf{49.73}     & \textbf{14.81}     & \textbf{35.56}     & &  \textbf{51.38}     & \textbf{16.63}     & \textbf{33.32}     & & \textbf{81.84}     & \textbf{94.25}     & \textbf{75.51} \\ \hline
\multicolumn{11}{l}{$^{*}$ Existing tools are not able to apply the obfuscation to large benchmarks.} \\
\multicolumn{11}{l}{$^{**}$ Excluding non-ISCAS85 benchmarks due to scalability issues in CS and SLL locking schemes.} \\
\end{tabular}}

\end{table*}

In this section, runtime, overhead, and security analysis are provided when applying the proposed approach to multiple ISCAS85, EPFL, and CEP benchmarks to ensure that our proposed approach provides the necessary protection at an affordable overhead. We use both Synopsys Design Compiler and ABC compiler for synthesis, and we map the benchmark designs on a 250nm LEDA standard cell library. Both ModelSim and Synopsys VCS are used for simulation. The testing hardware used in the analysis consists of a 4.4GHz CPU with 6 cores and 12 threads, and 32GB of DDR4 RAM.  

\subsection{Scalability Analysis}

\begin{figure}[h]
\centering
\includegraphics[width=0.5\textwidth]{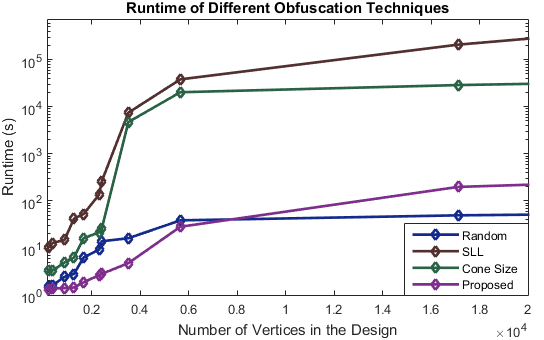}
\caption{Runtime vs number of vertices for our implementation of the existing and the proposed obfuscation approaches.}
\label{runtime}
\end{figure}

To confirm that our approach is scalable, the relation between runtime and circuit size (in terms of vertices) should not be exponential. Fig. \ref{runtime} shows the runtime comparison between our proposed process and some existing approaches such as random insertion, SLL, and cone-size-based insertion.

The runtime analysis shows that SLL and cone-size based approaches require an exponentially increasing runtime as the number of vertices increases, up to a point where some benchmarks cannot be practically obfuscated. On the other hand, the runtime analysis shows that our approach has a linear relationship between the number of vertices and the polynomial-time needed to complete the obfuscation process. This linear relation suggests that large IPs will be obfuscated without any scalability issues. We have also applied our approach to extremely large IPs (100,000+ gates), and the runtime results show that the approach is scalable. Table \ref{runtime_table} shows the runtime results when applying our approach to some large EPFL and CEP IPs.

\begin{table}[h]
\centering
\caption{Runtime Results for SARO applied to large IPs}
\label{runtime_table}
\resizebox{\linewidth}{!}{%
\begin{tabular}{ccccccc}
\hline
Benchmark & Number of Vertices & Number of Partitions &  Run Time (seconds)      \\
\hline
ALU         & 9264            & 458                  & 111.25            \\ 
mem\_ctrl   & 28759           & 1424                 & 277.05            \\
FIR         & 30797           & 1536                 & 282.35           \\ 
hypotenuse  & 214438          & 10571                & 7421.81            \\ 
AES192      & 302432          & 14734                & 9002.34           \\

\hline            
\end{tabular}}
\end{table}

\subsection{Overhead Analysis}

Although overhead constraints can be controllable when using our obfuscation tool, we aim to balance the quality of obfuscation and the cost in terms of area and power overheads. The tool always constrains the latency overhead not to exceed 5\% since latency dictates the maximum clock speed that can be used in the IP. 
Table \ref{overhead} shows the overhead analysis when applying our approach and setting the partition size to be 25 gates. It is clear from the overhead analysis that our obfuscation approach can maintain a low overhead value. The overall average for the tested benchmarks is 23.1\% for area overhead and 27.9\% for power overhead.

\begin{table}[h]
\centering
\caption{Overhead Results for SARO}
\label{overhead}
\resizebox{\linewidth}{!}{%
\begin{tabular}{ccccccc}
\hline
Benchmark &   Key Size &   \begin{tabular}{@{}c@{}}Number of\\ Partitions   \end{tabular}  & \begin{tabular}{@{}c@{}}Area \\  Overhead (\%)    \end{tabular} & \begin{tabular}{@{}c@{}}Power \\  Overhead (\%)    \end{tabular} & \begin{tabular}{@{}c@{}}Timing \\  Overhead (\%)    \end{tabular}\\
\hline
c432        & 32     &  8     & 28.79    & 39.62   &  4.61    \\
c499        & 32     &  11    & 31.01    & 29.08   &  5.00    \\
c880        & 32     &  14    & 32.89    & 47.40   &  4.50    \\
c1355       & 32     &  11    & 26.19    & 26.64   &  4.87    \\
c1908       & 32     &  12    & 27.50    & 27.52   &  5.12    \\
c2670       & 64     &  18    & 33.00    & 36.93   &  4.72    \\
c3540       & 64     &  39    & 44.52    & 65.84   &  5.07    \\
c5315       & 128    &  44    & 39.09    & 38.69   &  5.00    \\
c6288       & 256    &  141   & 34.42    & 39.29   &  4.66    \\
c7552       & 256    &  46    & 33.70    & 27.66   &  2.13    \\
ALU         & 256    &  458   & 31.21    & 27.71   &  4.78    \\
mem\_ctrl   & 512    &  1424  & 37.55    & 33.30   &  5.01    \\
FIR         & 512    &  1536  & 28.60    & 34.91   &  4.11    \\
hypotenuse  & 512    &  10571 & 29.47    & 27.33   &  4.81    \\
AES192      & 512    &  14734 & 24.95    & 39.07   &  5.01    \\

\hline  
\textbf{Average}      &    &          & \textbf{32.71 }         & \textbf{35.84  }    & \textbf{4.59  }    \\
\hline
\end{tabular}}
\end{table}

\subsection{Security Analysis}

To ensure that our approach provides the necessary IP protection, a thorough security assessment is applied to the obfuscated benchmarks. Various attacks have been applied to these benchmarks under different obfuscation settings. In this subsection, we will discuss the functional and structural alteration caused by SARO, as well as the security analysis against SAT, KSA, hill-climbing, SWEEP, desynthesis, and SAIL attacks.

\begin{figure*}[h]
\centering
\includegraphics[width=0.95\textwidth]{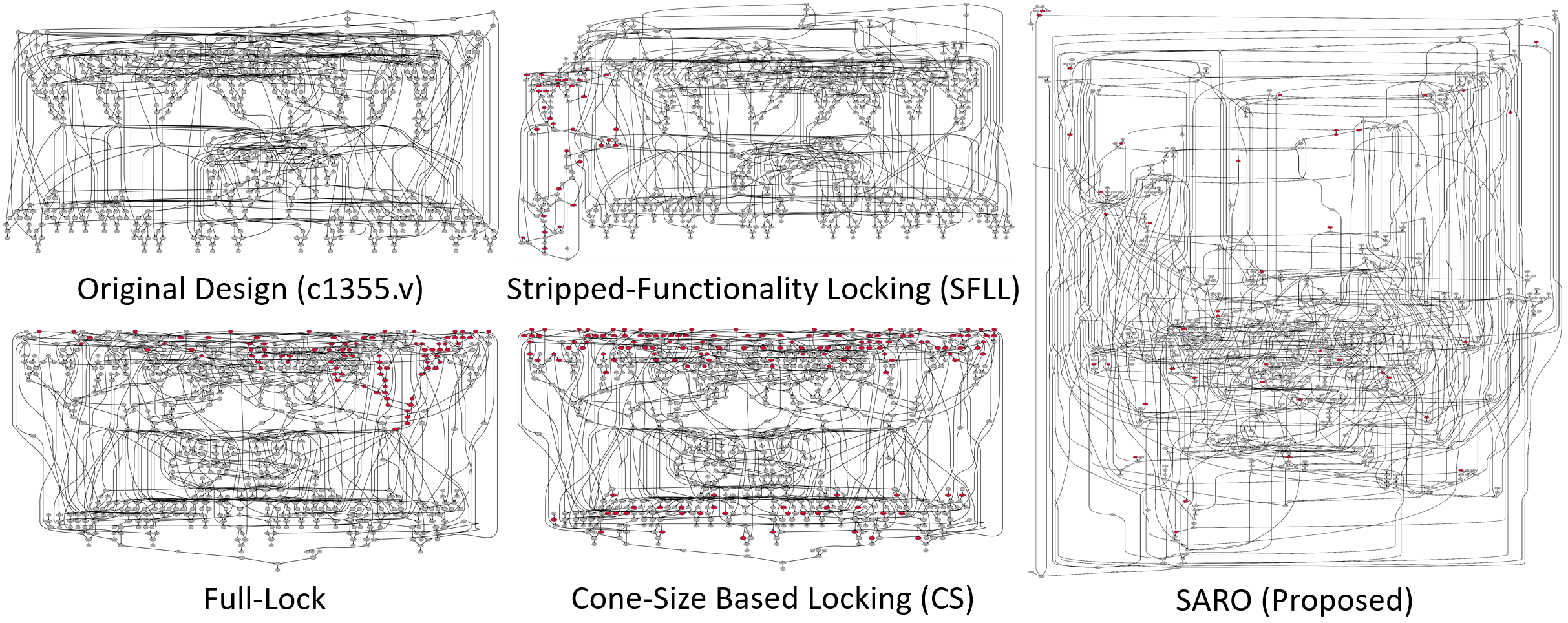}

\caption{A graphical overview of the c1355 benchmark is used to demonstrate the existing and proposed locking methods. The implemented methods are stripped-functionality logic locking \cite{yasin2017provably}, full-lock \cite{kamali2019full}, cone-size based locking \cite{lee2015improving}, and SARO. The obfuscation gates are highlighted in red. It is clear that in all existing methods key-entry nodes are clustered in one region of the circuit. However, SARO performs an obfuscation that distributes the key-entry nodes across the entire design and alters the graphical representation of the locked circuit.}
\label{schem_con}
\end{figure*}

\subsubsection{Functional and Structural Analysis}

Based on the results in Table \ref{struc}, it is clear that our approach causes the highest impact on the structure of the design. Getting high values for both formal verification and coverage index indicates that the obfuscation process causes a high level of functional alteration, while a high structural alteration is observed when getting a high scatter index value. 
The $T3_{metric}$ metric has been calculated for the existing and the proposed approaches and shown in Table \ref{s_metric}.

Since our approach focuses on local obfuscation, a strong structural transformation impact on the entire design is obtained. The optimization process can further alter the structural representation of the circuit. Hence, the design transformation solution offered by SARO would increase the complexity of pattern recognition attempts and hide all design signatures that might expose sensitive information about the system. One way to measure the structural differences between the original and the obfuscated designs is to analyze the proposed $T3_{metric}$ metric. The following analysis shows the formal verification test being applied to different benchmarks. Table \ref{s_metric} shows the functional and structural impact of the existing obfuscation techniques as well as SARO.

\begin{figure}[h]
\centering
\includegraphics[width=0.5\textwidth]{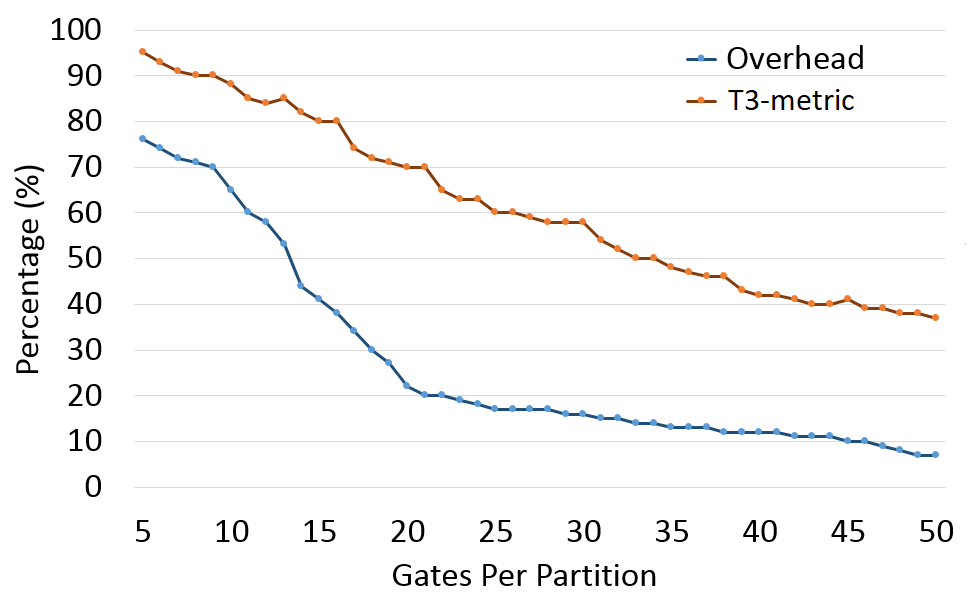}
\caption{The relation between structural alteration (represented as $T3_{metric}$) and the average overhead when changing the number of gates per partition. This analysis is based on c1355.}
\label{overs}
\end{figure}

The $T3_{metric}$ metric can be controlled based on the size of the partition. Smaller partitions would increase the $T3_{metric}$ score as a larger structural impact is caused to the obfuscated circuit. However, the smaller partitions are, the more overhead the design will have. Our obfuscation approach gives the user the ability to choose the size of the partitions, which can be useful when limited overhead is allowed. 
Fig. \ref{overs} shows the relation between the overhead (based on c1355) and $T3_{metric}$ when changing the number of gates per partition.

A visual comparison between the existing methods and SARO for the ISCAS85 c1355 benchmark is shown in Fig. \ref{schem_con}. The comparison shows that all existing methods insert key-entry nodes (highlighted in red) in a clustered fashion, where most of them are grouped in one area. However, key-entry nodes are well-distributed in SARO.

\begin{table}[h]
\centering
\caption{Functional and Structural Changes Analysis Using the Proposed $T3_{metric}$}
\label{s_metric}
\begin{tabular}{ccccc}
\hline
Benchmark & CS &  SLL &  RN &  \textbf{SARO} \\  \hline
c432  &         44.33    &    35.40    &  25.53    &   84.16      \\
c499  &         48.43    &    38.06    &  27.20    &   83.40       \\
c880  &         49.43    &    22.16    &  24.36    &   81.03      \\
c1355  &        49.86    &    17.73    &  44.93    &   75.50       \\
c1908  &        45.13    &    24.00    &  32.26    &   88.40       \\
c2670  &        55.16    &    29.33    &  39.33    &   83.96      \\
c3540  &        54.50    &    47.16    &  39.20    &   83.03      \\
c5315  &        63.73    &    38.36    &  34.16    &   83.83      \\
c6288  &        57.13    &    41.26    &  31.06    &   92.66      \\
c7552  &        68.16    &    40.16    &  39.70    &   82.66      \\
\hline 

\textbf{Average} & \textbf{53.59} & \textbf{33.36}    &  \textbf{33.77}     & \textbf{83.86} \\ \hline

\end{tabular}  
\end{table}

\subsubsection{Resiliency Against SAT Attack}

Due to the distributed SAT resistance, the obfuscated designs are well protected against SAT attack. The SAT attack runtime analysis results for existing obfuscation techniques and SARO are shown in Table \ref{table_sat}.

\begin{table}[h]
\centering
\caption{Runtime for SAT Attack to Break the Obfuscation}
\label{table_sat}
\begin{tabular}{ccccc}
\hline
Benchmark & CS &  SLL &  RN &  \textbf{SARO} \\  \hline
c432    & 1.17     &   2.25    &   1.21    &  timeout    \\ 
c499    & 2.22     &   2.21    &   1.17    &  timeout    \\ 
c880    & 4.35     &   1.11    &   3.29    &  timeout     \\ 
c1355   & 2.24     &   2.25    &   3.47    &  timeout     \\ 
c1908   & 4.34     &   2.22    &   3.35    &  timeout     \\ 
c2670   & 14.42     &   11.38  &   13.27    &  timeout     \\  
c3540   & 5.43     &   3.35    &   4.41    &  timeout     \\  
c5315   & 8.55     &   4.38    &   5.37    &  timeout     \\  
c6288   & timeout   &  timeout  & timeout    &  timeout     \\ 
c7552   & 15.59     &   8.41    &   8.37    &  timeout     \\ 

\hline 

\textbf{Average$^*$} & \textbf{6.47} & \textbf{4.17}  &  \textbf{4.87}  & \textbf{timeout} \\ \hline
\multicolumn{5}{l}{$^*$ Excluding the runtime of c6288.} \\

\end{tabular}  
\end{table}

All benchmarks using SARO obfuscation have not been broken through SAT until the set timeout (10 hours). While traditional obfuscation techniques have all been broken quickly. Even without the addition of the distributed SAT resistance, our approach is designed to obfuscate extremely large IPs (100,000+ gates), and the complexity of SAT grows as the design and the key-size increase. Moreover, large designs are more likely to contain SAT hard blocks that prevent the solver from converging. Hence, SAT attack should not cause a security threat when using our approach.

We have also tested the resiliency of SARO against AppSAT, which is an approximated SAT attack that converges when it finds a key that provides the correct functionality most of the time with minimal corruption that is often tolerable (e.g., 99\% of correct functionality) \cite{shamsi2017appsat}. When comparing our SAT resistance against existing methods, SARO can provide the necessary protection while maintaining high output corruptibility. Table \ref{SAT_ex_vs_saro} shows a brief comparison between SARO and existing SAT resistance techniques.

\begin{table}[h]
\centering
\caption{Summary of Existing and Proposed SAT Resistance Methods}
\label{SAT_ex_vs_saro}
\resizebox{\linewidth}{!}{%
\begin{tabular}{|c|c|c|c|c|c|c|}
\hline
   \begin{tabular}{@{}c@{}}\textbf{Features} \end{tabular}   & \textbf{ \begin{tabular}{@{}c@{}}AntiSAT\\ \cite{xie2016mitigating}\end{tabular}} & \textbf{ \begin{tabular}{@{}c@{}}SARLock\\ \cite{yasin2016sarlock}\end{tabular}}  & \textbf{ \begin{tabular}{@{}c@{}}SFLL\\ \cite{yasin2017provably}\end{tabular}} &\textbf{ \begin{tabular}{@{}c@{}}Full-Lock\\ \cite{kamali2019full}\end{tabular}}  &  \textbf{ \begin{tabular}{@{}c@{}}SARO\\ (Proposed)\end{tabular}}  \\  \hline
Approach & \multicolumn{2}{c|}{ \begin{tabular}{@{}c@{}}Reducing Keys\\ Eliminated per DIP \end{tabular} }&  \multicolumn{3}{c|}{\begin{tabular}{@{}c@{}}Increasing Complexity \\ of the SAT Solving \end{tabular} }\\ \hline
  \begin{tabular}{@{}c@{}}Output \\ Corruptibility  \end{tabular}    & Low     &   Low  &   High &   High &   High    \\ \hline
  \begin{tabular}{@{}c@{}}Number of Possible \\ SAT-Hard Functions   \end{tabular}   & Limited    &   Limited  &   Limited   &   Limited &  Unlimited      \\ \hline 
  \begin{tabular}{@{}c@{}}Structural \\ Signature   \end{tabular}   & Vulnerable     &   Vulnerable  &   Vulnerable  &   Vulnerable &   Hidden      \\ \hline
    \begin{tabular}{@{}c@{}}Average Overhead  \\ (Based on ISCAS85)\end{tabular}   & 15-40\%   &   10-22\%   &   10-30\%   &   50-400\%   &   24.1\%      \\ \hline

\end{tabular}   }
\end{table}

\subsubsection{Resiliency Against Hill-climbing Attacks}

Due to the key-size and the high output corruption of the obfuscated designs, hill-climbing attacks will not be easily performed. The accuracy of the initial guess of the key has to be very high before any observable functional behavior is achieved. The hill-climbing attack has been applied to an obfuscated c1908 benchmark using a 64-bit key; the attack algorithm has timed out (at 10-hours) without converging.

\subsubsection{Resiliency Against SWEEP and Desynthesis Attacks}

SWEEP and desynthesis attacks exploit the synthesis tool to determine whether the selected key is correct or wrong. However, SARO is designed to insert random dummy functions. The type and size of these functions vary, where it might be larger than the original functions in some instances, and smaller in others. This property reduces the accuracy of the attacks, where no common features can be extracted when the guessed key is correct. We have implemented the attacks on a SARO obfuscated c1908 benchmark, and the accuracy of the guessed key does not exceed 50\% in all the attempts that we have performed.
The attack accuracy analysis results for existing obfuscation techniques (mux-based with randomly selected dummy routes) and SARO are shown in Table \ref{table_sweep}.

\begin{table}[h]
\centering
\caption{Attack accuracy for SWEEP attack applied to existing techniques and SARO}
\label{table_sweep}
\begin{tabular}{ccccc}
\hline
Benchmark & CS &  SLL &  RN &  \textbf{SARO} \\  \hline
c432    & 64.0  &  58.0  &   75.0  &   0.0     \\ 
c499    & 75.0  &  88.2  &   72.5  &   5.0    \\ 
c880    & 88.2  &  100  &   88.2  &    0.0   \\ 
c1355   & 95.0  &  92.5  &   72.5  &   5.0    \\ 
c1908   & 48.5  &  66.5  &   32.6  &   2.5    \\ 
c2670   & 72.5  &  67.5  &   68.0  &   0.0    \\  
c3540   & 98.0  &  70.0  &   78.5  &   0.0    \\  
c5315   & 100   &  80.8  &   86.0  &   8.0    \\  
c6288   & 92.5  &  82.5  &   88.2  &   0.0    \\ 
c7552   & 78.6  &  64.0  &   77.5  &   2.5    \\ 

\hline 

\textbf{Average} & \textbf{81.2} & \textbf{77.0}  &  \textbf{73.9}  & \textbf{2.3} \\ \hline

\end{tabular}  
\end{table}

\section{Conclusion}

We have proposed SARO, a novel approach that applies a strong obfuscation that not only locks the design but also hides any structural signatures that might help attackers gain information about the system.
Additionally, the proposed approach applies a design modification process that is randomized in different design aspects, which significantly reduces the accuracy of structural analysis attacks that are based on machine learning and pattern recognition. We introduced a distributed attack resistance scheme that, unlike existing approaches, provides a well-hidden and hard-to-remove protection against all major attacks, while maintaining high output corruption when the wrong obfuscation key is applied. 
Moreover, the approach we propose proves to be very scalable compared to other existing techniques; runtime analysis shows that our approach is suitable for extremely large IPs. The tool applies the obfuscation technique that uses a customizable key size, and manageable latency, power, and area overheads. We have also introduced the $T3_{metric}$, which is an evaluation metric that quantifies the structural and functional design transformation level caused by the obfuscation process. Security assessment shows that the obfuscated circuits are protected against SAT, KSA, hill-climbing, and SWEEP. Further investigation should focus on incorporating the proposed tool to apply sequential obfuscation schemes.

\bibliographystyle{IEEEtran.bst}
\bibliography{IEEEabrv,bibliography}

\end{document}